\documentstyle[aps,preprint,epsfig]{revtex}
\tightenlines

\begin{document}

\title{ Statistical Mechanics of Vibration-Induced Compaction of Powders}
\author{S.F.Edwards and D.V.Grinev}
\address{Polymers and Colloids Group, Cavendish Laboratory, University
of Cambridge, Madingley Road, Cambridge CB3 OHE, UK}

\maketitle

\begin{abstract}
 We propose a theory which describes the density relaxation of
loosely packed, cohesionless granular material under mechanical tapping.
Using the compactivity concept we develop a formalism of statistical
mechanics
 which allows us to calculate the density of a powder as a function of
time and compactivity. A simple fluctuation-dissipation relation which
relates compactivity to the amplitude and frequency of a tapping is proposed.
Experimental data of E.R.Nowak {\it et al.} [{\it Powder Technology} {\bf 94}, 79 (1997) ] show how density
of initially deposited
in a fluffy state powder evolves
under carefully controlled tapping towards a random close packing (RCP)
density.
Ramping the vibration amplitude repeatedly up and back down again reveals the
existence of reversible and irreversible branches in the response.
In the framework of our approach the reversible branch (along which the RCP
density is obtained) corresponds to the
steady state solution of the Fokker-Planck equation whereas the irreversible
one is represented by a superposition of "excited states" eigenfunctions.
These two regimes of response are analyzed theoretically and a qualitative
explanation of the hysteresis curve is offered.
\end{abstract}
\vspace{2cm}
\noindent PACS numbers 81.05.R,05.40.,03.20.
\newpage

\section{Introduction}

There is an increasing interest in applying the methods of statistical
mechanics and  kinetic theory to granular materials where processes
are dominated by geometrical constraints and friction, and initially posses
a memory of sample's preparation \cite{Mehta,Jaeger}.

In this paper we propose an analytical approach which gives a qualitative
explanation of experimental data obtained by Nowak {\it et al.} \cite{Nowak}. They have shown that external vibrations lead to a slow, approach of the packing density to a final steady-state
value. Depending on the initial conditions and the magnitude of the vibration
acceleration, the system can either reversibly move between steady-state
densities or can become irreversibly trapped into metastable states that is
the rate of compaction and the final density depend sensitively on the
history of vibration intensities that the system experiences (see Fig.1).

A granular material is a system with a large number of individual
grains and therefore it has a huge number of degrees of freedom.
Grains interact with each other via contact forces which are determined by
friction, gravitational loading and amplitude of an exernal force
if the system is perturbed. Therefore one needs
to invent a formalism that would allow us to calculate macroscopic averages
in terms of microscopic (i.e. of individual grains) properties of the system.
If we assume that it may be characterised by a small number of
parameters
(e.g. analogous to temperature ) and that this system has properties
which are
reproducible given the same set of extensive operations (i.e. operations
acting upon the system as a whole rather than upon individual grains)
then we may apply the ideas of statistical averaging over the ensemble of
configurations to granular
systems \cite{Edwards}.

In the present paper we consider the simplest model of a granular material
by  introducing the volume function $W$ and assume  the simplest case that
all configurations of a given
 volume are equally probable; in many cases  the mechanism of
deposition will leave a history in the configuration but this will not
be considered here. $W$ will depend on the coordinates of the grains and their orientations and is the analogue of a Hamiltonian.
Averaging over all the possible configurations of the grains in
real space gives us a configurational statistical ensemble
describing the random packing of grains. Since we are assuming that we are dealing with a system whose constituents are hard (i.e. impenetrable) we have to include some account of this in our formalism in order to reduce the number of possible configurations the system may occupy. Also for a packing which is stable under applied force we must consider the configurations restricting the number of possible volume states that the system may occupy to be only those configurations which are stable. Also grains cannot overlap and this condition produces very strong constraints (frustration) on their relative positions. This implies that all grains have to be in contact with their nearest neighbours. Of course in the real powder the topological defects can exist such as vacancies, voids or arches. But as these will be a subject of a future paper we do not consider them here. 
Thus we have a "microcanonical" probability distribution \cite{Edwards}:

\begin{equation}\label{frank1}
 P=e ^{-\frac{S}{\lambda}}\delta(V-W)\,\,\Theta(contacts)
\end{equation}

\begin{equation}\label{frank2}
e ^{\frac{S}{\lambda}}=\int \delta(V-W)\,\Theta(contacts)\,\mathrm{d}(all\,degrees\,of\,freedom)
\end{equation}

where we define $\Theta$ as:

\[ \Theta(contacts) = \left\{ \begin{array}{ll}
                     1 & \mbox{if $z\geq z_{m}$} \\
                     0 & \mbox{if $z<z_{m}$}
                     \end{array}
                       \right. \]

where $z_{m}$ is the minimal coordination number of a grain \cite{Ball}. We have to introduce $\Theta$ because we consider the stable isotropic and homogeneous packings.
Just as in conventional statistical mechanics with microcanonical distribution:

\begin{equation}\label{frank3} 
P=e ^{-\frac{S}{k}}\delta(E-H)
\end{equation}

and temperature:

\begin{equation}\label{frank4} 
 T=\frac{\partial E}{\partial S}
\end{equation}

 we can define the analogue of temperature as:

\begin{equation}\label{frank5} 
 X=\frac{\partial V}{\partial S}.
\end{equation}

 This fundamental parameter is called compactivity \cite{Edwards}. It characterises
the packing of a granular material and may be interpreted as being
characteristic of the number of ways it is
 possible to arrange the grains in the system into volume $\Delta V$
such that the 
 disorder is $\Delta S$. Consequently the two limits of $X$ are $0$ and
$\infty$,
 corresponding to the most and least compact stable arrangements. This is clearly a valid parameter for sufficiently dense powders because one can
in principle calculate
the configurational entropy of an arrangement of grains and therefore derive
the compactivity from the basic definition \cite{compactivity}. One can expect despite the strong constraints resulting from the stability conditions, the number of packings to grow exponentially with the volume of a sample and the configurational entropy defined as a logarithm of this number is extensive. 

As usual it is more convenient to introduce the canonical probability distribution:

\begin{equation}\label{frank6}
 P=e ^{\frac{Y-W}{\lambda X}},
\end{equation}
where $\lambda$ is a constant which gives the entropy the dimension of
volume,
 $Y$ we call the effective volume, it is the analogue of the free energy:
\begin{equation}\label{frank7}
 e ^{-\frac{Y}{\lambda X}}=\int \,e ^{-\frac{W(\mu)}{\lambda X}}\,\mathrm{d (all)}, \qquad
 V=Y-X \frac{\partial Y}{\partial X}.
\end{equation}

To illustrate this theory consider the simplest example of a $W$, the analogue of Bragg-Williams approximation \cite{Edwards}: each grain has neighbours touching it with a certain coordination and angular direction. In order to set up an analogy with the statistical mechanics of alloys we assume that each grain has a certain property, which defines the ``interaction'' with its nearest neighbours. Taking the coordination number of a grain as such a property and assuming that there are just two types of coordination $z_{0}$ and $z_{1}$ we assign a volume $v_{i}$ to any grain with $z_{i}$ coordination number.
Thus we write the volume function as:

\begin{equation}\label{frank8}
 W=n_{0}v_{0} + (N-n_{0})v_{1}
\end{equation}

where N is the number of grains in the system, $n_{i}$ is the number of grains with the coordination number $z_{i}$ and $N=n_{0}+n_{1}$.
The simple calculation of $Y$ and $V$ gives us \cite{Edwards}:

\begin{equation}\label{frank9}
 Y=N \frac{(v_{0}+v_{1})}{2}-N \lambda X \ln2cosh\frac{(v_{0}-v_{1})}{\lambda X}
\end{equation}

\begin{equation}\label{frank10}
V=N \frac{(v_{0}+v_{1})}{2} + N \frac{(v_{0}-v_{1})}{2}tanh\frac{(v_{0}-v_{1})}{\lambda X}).
\end{equation}

Thus we have two limits: $V=Nv_{0}$, when $X \rightarrow 0$ and
$V=N(v_{0}+v_{1})/2$
 when $X \rightarrow \infty$ ($N$ is a number of grains).
Note that the maximum $V$ is not $Nv_{1}$ just as in the thermal
system (say a spin in a magnetic field) with two energy
 levels $E_{0}$ and $E_{1}$ one has $E=E_{0}$ when $T\rightarrow 0$ and
$E=(E_{0}+E_{1})/2$ when $T\rightarrow \infty$.

\section{ "Two-volumes" model: solution of the Fokker-Planck equation}

We consider the rigid grains powder dominated by friction deposited in a
container which will be shaken or tapped(in order to
consider the simplest case we ignore  other possible
interactions e.g. cohesion and do not distinguish between the grain-grain interactions in the bulk and those on the boundaries).
We assume that most of the particles in the bulk do not acquire any
 non ephemeral kinetic energy i.e. the change of a certain
configuration occurs due to continuous and cooperative rearrangement of
a free volume between the neighbouring grains. Any such powder will have a remembered
history of deposition and in particular
 can have non-trivial stress patterns, but we will confine the analysis
of this paper to systems
 with homogeneous stress which will permit us to ignore it.
The fundamental assumption is that under shaking a powder can return to
a well defined state,
independent of its starting condition. Thus in the simplest system, a
homogeneous powder, the
 density characterises the state.

It is sensible to seek the simplest
algebraic model for our calculation and to this end since the orientation of the grain must have at least two degrees of freedom, say $\mu_{1}$ and $\mu_{2}$, our volume function is:

\begin{equation}\label{frank11}
 W=v_{0}+(v_{1}-v_{0})(\mu_{1}^2+\mu_{2}^2)
\end{equation}

implying a two-dimensional picture (see Fig.2). When $\mu=0$
we have $W=v_{0}$ then the grain is ``well oriented'' which means that a free volume
 is minimal and  when $\mu=1$ and $W=v_{1}$ then the grain is ``not well oriented'' (free volume
is maximal).
It is a self-consistent approximation since the parameters $v_{0}$ and $v_{1}$ are the average volumes of the grain in the presence of other grains. 
In general we can write:

\begin{equation}\label{frank12}
e ^{-\frac{Y}{\lambda X}}=\int \, w(\mu)\,e ^{-\frac{W(\mu)}{\lambda X}}\,\mathrm{d}\mu
\end{equation}
where $w(\mu)$ is the weight factor attached to $\mu$.
From (\ref{frank7}) we derive $Y$ and $V$:

\begin{equation}\label{frank13}
 Y=N v_{0} - N \lambda X \ln  \Big\{ \frac{\lambda X}{v_{1} -
v_{0}}(1 - e ^{-\frac{v_{1}-v_{0}}{\lambda X}})\Big\}
\end{equation}

\begin{equation}\label{frank14}
V=N\big(v_{0} + \lambda X\big) - \frac{N(v_{1}-v_{0})}{e
^{\frac{v_{1}-v_{0}}{\lambda X}} - 1}.
\end{equation}

Thus we have the same limits as for volume function (\ref{frank8}): $V=Nv_{0}$, when $X \rightarrow 0$ and
$V=N(v_{0}+v_{1})/2$
 when $X \rightarrow \infty$.

 The main physical idea of our approach is the following:
all grains in the bulk experience the external perturbation as a
random force with zero correlation time so that the process of compaction can
be seen as the Ornstein-Uhlenbeck process for  the degrees of freedom $\mu_{i},\,i=1,2$ \cite{Risken}. Therefore we write the Langevin equation:
\begin{equation}\label{frank15}
  \frac{d \mu_{i}}{dt} + \frac{1}{\nu}\frac{\partial W}{\partial \mu_{i}}\,=\,\sqrt{D}\,f_{i}(t)
\end{equation}
where $\langle f_{i}(t)f_{j}(t')\rangle=2\delta_{ij}\delta(t-t')$ and $\nu$ characterises the frictional resistance imposed on the grain by its nearest neighbours. The term $f_{i}(t)$ on the RHS of (\ref{frank15}) represents the random force generated by a tap.  
The terms "shaken" or "tapped" have been used above and we have to
make them more precise.
The derivation
gives the analogue of the Einstein relation that $\nu=(\lambda X)/D$. If
we identify $f$ with the amplitude of the force $a$ used in the tapping,
the natural way to make this dimensionless is to write the ``diffusion''
coefficient  as :

\begin{equation}\label{frank16}
D=\Big(\frac{a}{g}\Big)^{2}\,\frac{\nu\sigma^{2}}{v}
\end{equation}
That is we have a simplest guess for a fluctuation-dissipation relation:

\begin{equation}\label{frank17}
\lambda X=\Big(\frac{a}{g}\Big)^{2}\,\frac{\nu^{2}\sigma^{2}}{v}
\end{equation}
where $v$ is the volume of a grain, $\sigma$ the frequency
of a tap and $g$ the gravitational acceleration.
Use of the Langevin equation (\ref{frank11}) is of
course a crude simplification as it does not explicitly take into account
the presence of boundaries and topological constraints. Generally speaking one would have to use the integro-differential
Langevin  equation with the memory kernel:
\begin{equation}\label{frank18}
\frac{d\mu_{i}}{dt}\,+\,\int_{0}^{t}\,K(t-t')\,\,\mu_{i}(t')\,d\,t'\,=\,\sqrt{D}\,f_{i}(t),
\end{equation}
 as one sees in experiment that the final
density depends sensitively on the history of vibration intensities.
Clearly to solve  such an equation is not a trivial task although the
solution could give us the better understanding of many  interesting features
of granular compaction.
The problem of how to choose the  initial values of $\mu$ is in reality
the  deposition problem. We discuss it later.

The Langevin equation can be easily solved for $W$ quadratic in $\mu$:

\begin{equation}\label{frank19}
\mu_{i}(t)=\mu_{i}(0)e^{-\gamma t}\,+\,\sqrt{D}e^{-\gamma t}\,\int_{0}^{t}\,f(t')e^{\gamma t'}\mathrm{d}t'
\end{equation}
Averaging over the ensemble we get:
\begin{equation}\label{frank20}
\langle\mu_{i}(t)\rangle\,=\,\mu_{i}(0)e^{-\gamma t}
\end{equation}
where $\mu_{i}(0)=1$ is the initial value of $\mu_{i}, \gamma=2\frac{(v_{1}-v_{0})}{\nu}$
has the meaning of relaxation time of the degree of freedom $\mu$.
As $t\rightarrow\infty$ $\mu$ goes to $\mu_{f}=0$ which corresponds to the
random close packing limit. 
The Fokker-Planck equation seems to be quite generic in modelling the response of granular materials to an externally applied shear rate \cite{shear} although in that problem it is more convenient to use the volume ``Hamiltonian'' $W$ as a function of the coordination number of each grain.  
The standard treatment of the Langevin equation (\ref{frank15}) is to use it to derive the
Fokker-Planck equation:.   
\begin{equation}\label{frank21}
 \frac{\partial P}{\partial t} =\Big( D_{ij}\frac{\partial^{2}}{\partial
\mu_{i}\partial \mu_{j}}+\gamma_{ij}\frac{\partial}{\partial\mu_{i}}\mu_{j} \Big)P=0
\end{equation}
where $D_{ij}=D\delta_{ij}$ and $\gamma_{ij}=\gamma\delta_{ij}$. 
 Equation (\ref{frank21}) can be solved explicitly. It has right- and left-hand
eigenfunctions $P_{n}$ and $Q_{n}$ and eigenvalues $\omega_{n}$ such
that:

\begin{equation}\label{frank22}
\omega_{n} P_{n}=\frac{\partial}{\partial \mu_{j}}\Big(D_{ij}\frac{\partial}
{\partial \mu_{i}} +\gamma_{ij}\mu_{j}\Big)P_{n}
\end{equation}

\begin{equation}\label{frank23}
\omega_{n} Q_{n}=\Big(-D_{ij}\frac{\partial}{\partial\mu_{i}} + \gamma_{ij}\mu_{j} \Big)\frac{\partial}{\partial\mu_{j}}Q_{n}
\end{equation}
or equivalently a Green function:
\begin{equation}\label{frank24}
G=\sum_{n}P_{n}(\mu)Q_{n}(\mu)e ^{-\omega_{n}t}.
\end{equation}
It follows that if we start with a non-equilibrium distribution:
\begin{equation}\label{frank25}
P^{(0)}(t=0)=\sum_{n=0}^{\infty}A_{n}P_{n}, \qquad A_{n}=\int
Q_{n}P^{(0)}\,\mathrm{d}\mu_{1} \,
\mathrm{d}\mu_{2}
\end{equation}
 and it will develop in time as:

\begin{equation}\label{frank26}
P^{(0)}(t)= A_{0}P_{0} + \sum_{n\neq 0}^{\infty}A_{n}P_{n}e
^{-\omega_{n}t}
\end{equation}
where $\int\, P^{(0)}\,\mathrm{d}\mu_{1}\mathrm{d}\mu_{2}\,=\,A_{0}$. This  coefficient is determined by a
 number of grains present in the powder, hence must be a constant.
The steady-state distribution function is:

\begin{equation}\label{frank27}
P^{(0)}(t\rightarrow \infty)=\frac{e ^{-\frac{(v_{1}-v_{0})(\mu^{2}_{1}+\mu^{2}_{2})}{\lambda X}}}{\int_{0}^{1}\,e ^{-\frac{(v_{1}-v_{0})(\mu^{2}_{1}+\mu^{2}_{2})}{\lambda X}}\,\mathrm{d}\mu_{1}\,\mathrm{d}\mu_{2}}
\end{equation}
The Fokker-Planck operator (\ref{frank21}) has a complete orthogonal set of eigenfunctions:

\begin{equation}\label{frank28}
P_{n} =H_{n}e ^{-\frac{(v_{1}-v_{0})(\mu^{2}_{1}+\mu^{2}_{2})}{\lambda X}} 
\end{equation}

where $H_{n}$ are Hermite polynomials and $\mu_{i} \in\, (0, \infty)$. In our case $\mu_{i} \in\,(0,1)$. One can avoid this mathematical difficulty, taking into account the crudety of our model and constructing the ``first excited state'':  $P_{2}=(a(\mu^{2}_{1}+\mu^{2}_{1})+b)\,e ^{-\frac{(v_{1}-v_{0})(\mu^{2}_{1}+\mu^{2}_{1})}{\lambda X}}$ orthogonal to the ground-state eigenfunction $P_{0}$. This eigenfunction describes the initial state of our system i.e. loosely packed deposited powder. Therefore it is easy to see the initial non-equilibrium distribution (\ref{frank26}) depends on how the the powder is deposited.   Constants $a$ and $b$ can be defined from the orthonormality relations. By using:
\begin{equation}\label{frank29}
P_{n}=Q_{n}\,P^{(0)}(t\rightarrow\infty),\quad Q_{0}=1
\end{equation}
and:

\begin{equation}\label{frank30}
\int_{0}^{1} Q_{2}\,\hat{L}_{FP} \,P_{2} \,\mathrm{d}\mu_{1}\,\mathrm{d}\mu_{2}=\omega_{2}
\end{equation}
one can easily verify that the eigenvalue $\omega_{2}$ (which corresponds to $P_{2}$ and gives us the decay rate of our nonequilibrium distribution) is a constant dimensionless number.      

Suppose now that deposition produces a
 highly improbable configuration, indeed
 the most improbable configuration whith: $\mu_{1}^{2}+\mu_{2}^{2}=2$ and the mean
volume
 function is $\bar{W}=2v_1-v_{0}\,\approx\,v_{1}$, where:

\begin{equation}\label{frank31}
\bar{W}(X,t)=\int P ^{(0)}(X,t)W \,\mathrm{d}\mu_{1}\mathrm{d}\mu_{2}.
\end{equation}

It is possible to imagine a state where all the grains are improbably
placed, i.e. where each
grain has its maximum volume $v_{1}$. In a thermal analogy this would
be like fully magnetised  magnetic array of spins where the magnetic field is suddenly reversed. Such a
system is highly unstable and equilibrium statistical
 mechanics does not cover this case at all. It will thermalize consuming
the very high energy whilst
establishing the appropriate temperature. Powders however are dominated
by friction, so if one could put
 together a powder where the grains were placed in high volume
configuration, it will just sit there until
 shaken; when shaken it will find its way to the distribution
(\ref{frank6}). It is possible to identify
physical states of the powder  with characteristic values of volume in
our model. The value $V=Nv_{1}$ corresponds to the ``deposited''
powder, i.e. the powder is put into the most unstable condition
possible, but friction holds it.   When $V=Nv_{0}$ the powder is
shaken into closest packing. The intermediate value of
$V=(v_{0}+v_{1})/2$ corresponds to the minimum density of the
reversible curve. Thus we can offer an interpretation of three values of density presented in the  experimental data \cite{Nowak}.

The general solution of the Fokker-Planck equation (\ref{frank21}) goes
to its
 steady-state value when $t\rightarrow \infty$ so we can expect
$\bar{W}(X,t)$
to diminish (as the amplitude of tapping increases) until one reaches
the
 steady-state value $\bar{W}(X)$. The formula (\ref{frank14}) can be
obtained
 using (\ref{frank31}) when $t\rightarrow \infty$ and represents a
reversible
 curve in experimental data of \cite{Nowak}: altering $a$ moves one
along the curve
 $\rho=\frac{v}{\bar{W}(X)}=\rho(a)$.
 We can identify time with the number of taps, so wherever we start with
any
 initial $\rho_{(0)}$ and $a$, successive tapping takes one to
reversible
curve $\rho(a)$. Or, if one decides on a certain number of taps,
$t\neq\infty$
, one will traverse a curve $\rho_{t}(a)$, where
$\rho_{\infty}(a)=\rho(a)$.
Notice that the simple result lies within the crudety of our
 model. The general problem will not allow us to think of
$X$ as
 $X(a)$ independent of the development of the system. The thermal
analogy is
 this: if the Brownian motion in an ensemble of particles is controlled
by a random
 force $f$ which is defined in terms of its amplitude and time profile,
this
 random force defines the temperature in the system. Our problem is like
a
 magnetic system  where magnetic dipoles are affected by a constant
magnetic
field, being random at high temperature , and increasingly oriented by
the
 external field as the temperature falls.

\section{Discussion}

The physical picture presented in section 2  is consistent with everyday
 knowledge of granular materials: when poured they take up a low density
but
 when shaken settle down, unless shaken violently when they return to
low
 density. These effects are much more pronounced in systems with
irregularly
 shaped grains then with fairly smooth uniform spheres, indeed the more
irregular a grain is, the more the  discussion above describes big
differences
 between $\rho_{(0)}$ and $\rho$. The experimental data of \cite{Nowak} show
 the  packing density dependence on parameter $\Gamma=a/g$ for a fixed number of taps. A loosely
packed bead assembly first undergoes irreversible compaction
corresponding to the lower branch of $\rho(\Gamma)$. The
settling behavior becomes reversible only once a characteristic
acceleration has been exceeded.
 Our theory gives three points $\rho(X=0), \rho(X=\infty)$ and
$\rho(t=0)$ which are in the ratio: $v_{0}^{-1},\,\frac{2}{(v_{0}+v_{1})},\,
v_{1}^{-1}$ and these are in reasonable agreement with experimental data: $\rho(X=0)=\frac{1}{v_{0}}\approx 0.64,\, \rho_{0}=\frac{1}{v_{1}}\approx 0.58$ and $\rho(X=\infty)=\frac{2}{(v_{0}+v_{1})}\approx 0.62$. Another important issue is the validity of the  compactivity concept for a ``fluffy'' but still mechanically stable granular arrays e.g. for those composed of spheres with $\rho \le 0.58$. In our theory $\rho(X=\infty)$ corresponds to the beginning of the reversible branch (see Fig.1) and using our analogy with a
 magnetic system is analogous to dipoles at a high temperature. The irreversible branch has an analogue in the behaviour of the magnetic system where initially the dipoles are strongly aligned with an external field but this field is then flipped to the opposite direction. 

The fluffy powder is a very complicated object as it has plenty of topological defects and stress arches. Throughout the paper we assumed that our granular array is spatially homogeneous which is the case for densities of the reversible curve.  However this is a very subtle problem which will be a subject of a future paper.  It is a difficult problem to decide whether
embarking on a vast amount of algebraic work that a superior mode would
entail is worthwhile. But our simple model is quite physical and can be extended when experiments would justify the work. 

 A final point is that we find the lower (irreversible)
curve build up to the upper (reversible) curve exponentially in time:

\begin{equation}\label{frank32}
V(t)= V_{\mathrm{\footnotesize initial}}\,e ^{-\omega
t}+V_{\mathrm{\footnotesize final}}(1-e ^{-\omega t})
\end{equation}

while one can  expect the logarithmic in time approach to the steady state
density e.g. the Vogel-Fulcher type curve which is typical
of disordered thermal systems such as spin and structural glasses \cite{glass}

\begin{equation}\label{frank33}
V(t) =V_{f}+(V_{i}-V_{f})e ^{-\omega t} + Tt^{-\epsilon}
\end{equation}
where $\epsilon$ is large.

 In fact, identifying $t$ with the number of taps $n$, the law seems to
be even slower at $(\ln t) ^{-1}$. Our simple analysis is clearly inadequate
to
obtain such a result which is quite outside the straightforward  method of expansion in the present set of eigenfunctions. However there is an argument by de Gennes \cite{de Gennes} which argues that a Poisson distribution
can provide this logarithmic behaviour.

\section{Acknowledgement}
 
This research was carried out as part of the DTI Colloid Technology Link
Project supported by Unilever, ICI, Zeneca and Schlumberger.
S.F.E. acknowledges an Emeritus Fellowship from Leverhulme Foundation.
D.V.G. acknowledges a Research Studentship from Shell (Amsterdam).

\newpage

\begin{figure}
\centerline{\psfig{figure=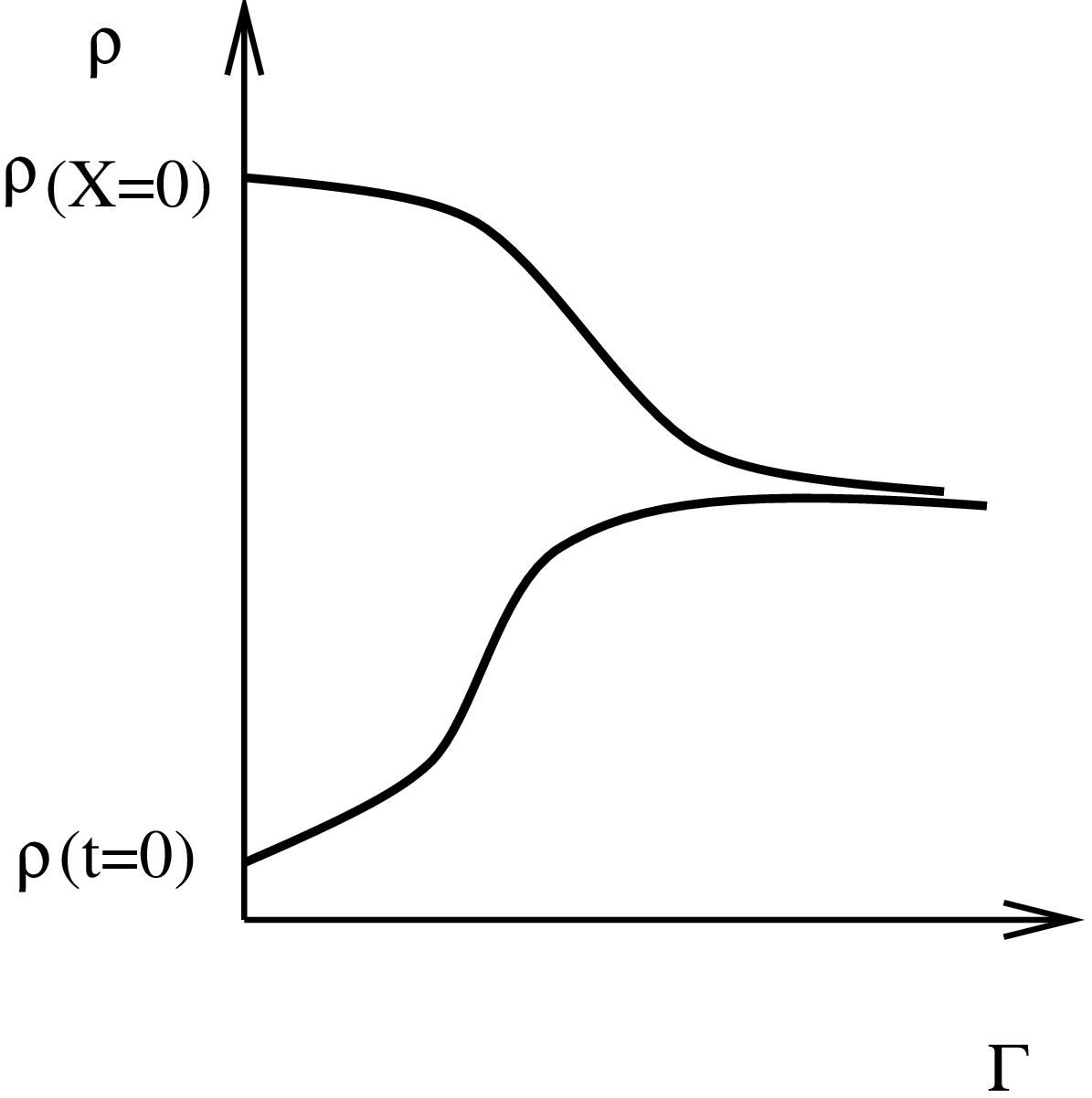,width=10cm,angle=-90}}
\vspace{2cm}
\caption{Dependence of the steady-state packing density on the tapping history (Nowak et al.). Experimental values of density packing fraction are in the following correspondence with  model parameters: $\rho(X=0)=\frac{1}{v_{0}}\approx 0.64, \rho(t=0)=\rho_{0}=\frac{1}{v_{1}}\approx 0.58$ and $\rho(X=\infty)=\frac{2}{(v_{0}+v_{1})}\approx 0.62$.}
\end{figure}

\begin{figure}
\centerline{\psfig{figure=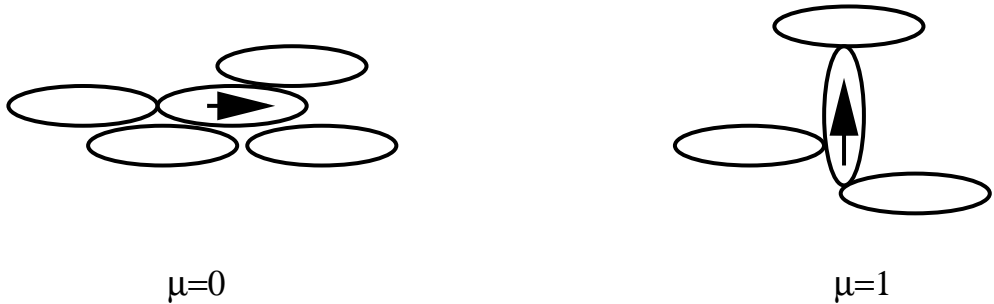,width=4cm,angle=-90}}
\vspace{2cm}
\caption{Graphical representation for the limit values of the degree of freedom $\mu$ in 2-D  }
\end{figure}


\begin{thebibliography}{99}

\bibitem{Mehta}{\it Granular Matter: An
Interdisciplinary Approach}, edited by A.Mehta (Springer-Verlag, New-York, 1993).
\bibitem{Jaeger} H.M.Jaeger, S.R.Nagel, and R.P.Behringer, Rev.Mod.Phys. {\bf 68}, 1259 (1996).
\bibitem{Nowak}E.R.Nowak, J.B.Knight, M.Povinelli, H.M.Jaeger
and S.R.Nagel, Powder Technology {\bf 94}, 79 (1997).
\bibitem{Edwards}S.F.Edwards, in {\it Current Trends in the
    Physics of Materials}, (Italian Physical Society and North Holland,
  Amsterdam, 1990), S.F.Edwards  and R.B.S.Oakeshott,
Physica A {\bf 157}, 1080 (1989), S.F.Edwards and C.C.Mounfield,
    Physica A {\bf 210}, 279 (1994).
\bibitem{Ball} R.C.Ball and S.F.Edwards, {\it to be published}.
\bibitem{compactivity} R.Monasson, O.Pouliquen, Physica A {\bf 236}, 395 (1997).
\bibitem{Risken} H.Risken, {\it The Fokker-Planck Equation}, (Springer-Verlag, New-York, 1990).
\bibitem{shear} A.Higgins and S.F.Edwards, Physica A {\bf 189}, 127 (1992).
\bibitem{glass} M.Mezard, G.Parisi, and M.A.Virasoro, {\it Spin Glasses and Beyond}, (World Scientific, Singapore, 1987) .
\bibitem{de Gennes} T. Boutreaux and P.G.de Gennes, {\it preprint}.

\end{thebibliography}
\end{document}